
\def\solar{\odot}
\def\lsun{\hbox{$L_{\odot}$}\tcar}
\newcount\eqnumber
\eqnumber=1
\def\chaphead{}
\def\new{\hbox{(\chaphead\the\eqnumber}\global\advance\eqnumber by 1}
\def\ref#1{\advance\eqnumber by -#1 (\chaphead\the\eqnumber
     \advance\eqnumber by #1 }
\def\first{\hbox{(\chaphead\the\eqnumber{a}}\global\advance\eqnumber
by 1}
\def\last{\advance\eqnumber by -1
\hbox{(\chaphead\the\eqnumber}\advance
     \eqnumber by 1}
\def\eq#1{\advance\eqnumber by -#1 equation (\chaphead\the\eqnumber
     \advance\eqnumber by #1}
\def\eqnam#1{\xdef#1{\chaphead\the\eqnumber}}
\def\eqt#1{Eq.~({{#1}})}

\def\lta{\mathrel{\spose{\lower 3pt\hbox{$\mathchar"218$}}
     \raise 2.0pt\hbox{$\mathchar"13C$}}}
\def\gta{\mathrel{\spose{\lower 3pt\hbox{$\mathchar"218$}}
     \raise 2.0pt\hbox{$\mathchar"13E$}}}
\baselineskip=24pt
\overfullrule=0pt
\def\doublespace{\baselineskip=2\normalbaselineskip}
\def\Rf{\doublespace\parindent=0pt \medskip\hangindent=3pc
\hangafter=1 }
\def\testnextcar{\ifhmode\ifcat\next.\else\ \fi\fi}
\def\tcar{\futurelet\next\testnextcar}

\def\narrower{\advance\leftskip by 30pt \advance\rightskip by 30pt}
\centerline{g-modes and the Solar Neutrino Problem}
\bigskip
\medskip\centerline{JOHN N. BAHCALL}
\centerline{Institute for Advanced Study, School of
Natural Sciences,
Princeton, New Jersey 08540}
\smallskip
\centerline{PAWAN KUMAR}
\smallskip
\centerline{Massachusetts Institute of Technology, Dept.
of Physics,
Cambridge, MA 02139}
\bigskip
\centerline{ABSTRACT}
\medskip
{\smallskip\narrower
We show that low-order g-modes
with large enough
amplitudes to affect significantly the solar neutrino fluxes would produce
surface velocities that are $10^4$ times larger than the observed
upper limits and hence are ruled out by existing data. We also
demonstrate that any large-amplitude,
short-period oscillations that grow on a Kelvin-Helmholtz time scale
will require, to affect solar neutrino fluxes,
a large amount of energy (for g-modes,
$10^9$ times the energy in the observed $p-$ mode oscillations) and
a tiny amount of dissipation (for g-modes, $10^{-8}$ the
dissipation rate of the $p$-modes).
\smallskip}
\bigskip
\noindent
\noindent
\centerline{1.  INTRODUCTION}
\medskip
Press (1981) and Press \& Rybicki (1981) have explored the possibility
that hydrodynamic phenomena involving g-modes  could change the
conventional
quasistatic picture of main-sequence evolution.
The particular process that Press and Rybicki investigated
involved the hydrodynamic coupling, by g-modes of high radial
wavenumbers, between turbulent fluid motions at the base of the
convective zone and shear fluid motions closer to the solar core.
They concluded that the coupling was not sufficient to drive large
time-dependent phenomena in the region in which the neutrinos are
produced.  A later study by Spruit (1987) suggested that the
coupling is probably even smaller than estimated by Press (1981).

The effect of low order g-modes on the solar interior has been
considered by many authors following a series of stimulating initial
discussions by Dilke \& Gough (1972), Ulrich (1974), and
Christensen-Dalsgaard, Dilke, \& Gough (1974) (for recent references,
see Merryfield, Toomre, \& Gough 1991) of a possible instability
associated with the nuclear burning of $^3$He.  Most recently, Gough
(1991) has considered a schematic model in which hypothesized very
large amplitude ($\delta T/T ~\sim~0.1$) temperature fluctuations in
the center of the sun are driven by low order, low degree (small
radial
and angular wave numbers) g-mode oscillations.  Gough (1991) and De
R\'ujula \& Glashow (1992) have suggested that the hypothesized
temperature fluctuations could lower significantly the neutrino fluxes
calculated in the standard solar model and produce an observable time
dependence in the $^8$B neutrino flux that would be detected in the
second generation solar neutrino detectors now under construction.  As
emphasized by Gough (1991) and by De R\'ujula \& Glashow (1992), it is
important to evaluate the possible effects of g-modes on the solar
neutrino fluxes since the existing conflict between predictions made
using the standard solar model and the results of the four operating
solar neutrino experiments have important implications for astronomy
and physics.

In this paper we concentrate primarily on the constraints on
large amplitude g-modes, whose existence has been hypothesized by Gough
(1991)
and De R\'ujula \& Glashow (1992) to resolve the solar neutrino
problem.
However, in \S\ 2, we first describe some order-of-magnitude
considerations
that show why it is difficult for low order g-modes, or any
similar short-period
oscillations that grows on a Kelvin-Helmholtz time scale, to
influence significantly the
neutrino fluxes.
In \S 3, we present the results of detailed calculations of the g-mode
eigenfunctions that show that temperature fluctuations large enough to
significantly affect the solar neutrino fluxes would produce
characteristic velocity fluctuations on the solar surface that are
$10^4$ times larger than the observed upper limits.
We summarize our conclusions in \S 4.
\vfill\eject
\noindent
\centerline{2.  ORDER--OF--MAGNITUDE CONSIDERATIONS}
\medskip
\medskip
\noindent

In order to produce temperature fluctuations that significantly affect
the neutrino fluxes, the hypothesized g-modes must have large
amplitudes. In the linear regime, the g-mode amplitude  is
approximately proportional to the fractional temperature change,
$\delta T/T$.  Gough (1991) estimates that a periodic fluctuation in
the central temperature of order $\delta T/T ~\sim~0.1$ is required to
solve the solar neutrino problem.  This result is consistent with the
fact that different central temperatures with a range of order $\delta
T_{\rm central}/T_{central} \sim~~0.02$ yield calculated neutrino
fluxes within the quoted standard theoretical uncertainties (Bahcall
 \& Ulrich 1988).

The energy content of low-order g-modes is approximately

\eqnam{\energygmodes}
$$
E_{\rm g} ~\sim~ M_{\rm core} V^2 ~\sim~ (\delta T/T)^2 M_{\rm core}
 R_{\rm core}^2 \tau_{\rm period}^{-2} ~~\sim~~(\delta T/T)^2  10^{45}
{\rm erg},
\eqno\new)
$$
where $V$ is fluid velocity associated with the mode, $M_{\rm core}
\simeq 0.2~M_\solar$ and $R_{\rm core} \sim 0.15~R_\solar$.
In deriving the
second part of the above
equation, we have made use of the fact that the radial wavelength of
low
order g-modes is $\sim R_{\rm core}$.  This energy content is much
larger than the energy in the well-observed and accurately-studied
solar p-modes
(Leibacher et al. 1985; Toomre 1986; Brown, Mihalas, \& Rhodes 1986;
Libbrecht \& Woodard 1991; Christensen-Dalsgaard \& Berthomieu 1991;
Guzik \& Cox 1992).  Using the result for the total energy in
approximately ten million solar p-modes (Libbrecht \& Woodard, 1991)
one finds that

\eqnam{\ratiogtop}
$$
{ {E_{\rm g}} \over  {E_{\rm p}}} ~>~
(\delta T/T)^2  10^{11} .
\eqno\new)
$$
Thus, for the required $ (\delta T/T)~=~0.1$,
the energy content of the hypothesized g-modes is more than
$10^9$ times the energy content of the p-modes.

The huge amount of energy required in the large-amplitude
g-modes severely restricts the possible physical mechanisms for their
excitation. The source of this energy must be either the nuclear fuel
in the solar core or the turbulence in the convective zone. The energy
in g-modes due to convective excitation (e.g. Goldreich \& Keeley
1977,
Spruit 1987, and Goldreich \& Kumar 1988) is
estimated to be small -- on the order of the kinetic energy in
a turbulent eddy of turnover time equal to the mode period. For
g-modes of period about an hour, this convective energy is more
than ten orders of magnitude
smaller than required to significantly affect the solar neutrino
fluxes.

A number of authors have investigated a possible instability
associated
with the high temperature sensitivity of $^3$He burning in the solar
core (Dilke \& Gough 1972;
Rosenbluth \& Bahcall 1973; Christensen-Dalsgaard et al. 1974;
Ulrich 1974; Saio 1980 and references quoted therein). It should be
noted that the results appear to depend upon the choice of the solar
model used as well as the modeling of the interaction of oscillation
with radiation and turbulence near the solar surface.
Rosenbluth \& Bahcall (1973), using an accurate solar model and
the full detail of the nuclear reactions, found $\ell$=1 modes to be
stable.
All authors agree that if the instability is present, it is largest
for
modes of low spherical harmonic degree ($\ell$) and order $n$.  In
fact
most authors find that g-modes of degree 2 and higher are stable in
the
present sun (Boury et al. 1975, Shibahashi et al. 1975, and Saio
1980), which
is not surprising since the radiative damping increases with the
square of the
inverse wavelength or $\ell^2$, whereas driving is expected to be
roughly constant for low order g-modes.

Can low order g-modes attain the large amplitudes required to modify
the solar neutrino fluxes?  The following considerations suggest that
this is unlikely. The ratio of period to characteristic growth time,
$Q^{-1}$, for g-modes is

\eqnam{\QRatio}
$$
Q^{-1} ~=~ { {1~hr} \over {10^7y} } ~ \sim 10^{-11} .
\eqno\new)
$$
We have used in \eqt{\QRatio}
the fact that the characteristic growth time for low order g-modes
is of order the Kelvin-Helmholtz time (see, e.g.,
Christensen-Dalsgaard et al.
1974). Thus modes will be stabilized if more than one part in
$10^{11}$
of the energy is dissipated per cycle (Press 1986).
It is hard to imagine that such a dissipationless instability will be
found.  We note that the observed $p$-modes have a fractional
dissipational rate of $Q^{-1}~\sim~10^{-3}$, eight
orders of
magnitude larger than the Q-value of g-modes required to change the solar
neutrino
fluxes.

Press (1986) has suggested that the energy dissipation in the
convection zone associated with the interaction with convective
eddies will
be an important limiting process. In fact, according to our numerical
calculations, the dissipation time due to turbulent viscosity for
the $g_1$
mode (n=1) of $\ell$=1 is about a factor of 40 smaller than its
growth time of 10$^7$ years.  The turbulence dissipation rates for
$g_2$
and $g_3$ modes of $\ell$=1 are about a factor of four  greater than their
growth rates due to the $^3$He instability, and the dissipation
rates for
$\ell$=2 modes are about a factor of 30--100 greater than their growth
rates. We have calculated the turbulent dissipation rate, following
Goldreich and Keeley (1977), by assuming an effective eddy viscosity
($\lambda V_\lambda$, where $\lambda$ and $V_\lambda$ are the size and
the velocity of the largest turbulent eddies such
that $V_\lambda/\lambda$ is equal
to or greater than the mode frequency $\omega$), and using the
viscous stress
tensor for nonradial oscillations (see Landau  and Lifshitz, 1982).
Due to
uncertainties associated with the modeling of turbulent viscosity our
results are uncertain by a factor of a few, but we  think it
unlikely that
the results are off by a factor of 40 (for comparison the computed
linewidth of p-modes, around 3mHz, due to  turbulent dissipation is
within
a factor of 2 of their observed linewidths). Therefore $g_1$ modes
of low
degree, if subjected to an instability acting on Kelvin-Helmholtz time
scale, will be stabilized due to turbulence dissipation. The
dissipation
rates for modes of $\ell$=1 and order (n) greater than one are close
to
their calculated growth rates, and thus, for these modes, we can not
conclude
that the $^3$He instability in the core, if present, is
suppressed. We note
that several people have included the perturbation to the convective
flux in
their linear stability calculations (e.g. Boury  {\it et.al.} 1975,
and Saio
1980). However, we are not aware of any previous paper that  has
considered the
turbulence dissipation of g-modes, which we find to be an important
process. In the future, the interaction of convection
with oscillations  should be
calculated carefully by proponents of any instability.

Suppose, despite the difficulties, that the low degree g-modes
manage to
grow to large amplitudes. Could one hide these large amplitude waves
in
the solar core?  Gravity waves propagate only in a stably stratified
medium, and are evanescent in the outer third of the sun, which is
unstable to convection.  One might try to conceal large amplitude
g-modes inside the core by assuming that they will be dissipatively
damped outside the core and so will go undetected at the solar
surface. As is shown in the following section, if the dissipation
time,
$\tau_{\rm dissipative}$, is much longer than the mode period then
one would
expect to see surface velocities associated with the g-modes that
are much
larger than the current observational limits. If one
tries to avoid the observational limit by assuming $\tau_{\rm
dissipative}$
to be of order the mode period, then the effective luminosity due to
g-modes,
$L_{\rm effective, g}$, is much larger than the standard radiative
luminosity,
$
L_{\rm effective, ~g} ~=~
{ {E_{\rm g}} \over  {\tau_{\rm dissipative}}} \sim \left({\delta
T\over T}
  \right)^2 10^8 \lsun.
$
Intuitively, one would expect that
such a large perturbation from the standard solar model
would show up in some discrepancy between the standard model
and the many available observations of the sun.
In any event, the scenario represented by this huge luminosity
would not yield a self-consistent solution since the
hypothesized damping rate is much greater than the calculated growth
rates.

In the next section we present
results of the numerical calculation of solar g-mode eigenfunctions
which
show that the attenuation factor for low degree g-modes in the
convection zone is not large, and so they should be easy to detect at
the surface.  But, in the spirit of the order-of-magnitude estimates
of
this section, we note that the radial wavenumber for a low frequency
high
degree gravity wave, which is of course imaginary in the convection
zone, is roughly equal to the horizontal wavenumber or $\ell/r$.
Therefore, g-mode amplitude varies approximately as
$1/r^{\ell}$ in the convection zone, and so large $\ell$ g-modes will
have small amplitudes at the surface. However, amplitudes of low
degree modes at the surface are not much smaller than the value at the
bottom of the convection zone. Finally, note that the velocity
amplitude, $V_{\rm core}$, and the temperature fluctuation amplitude,
$(\delta T)_{\rm core}$, in the core region for the g-modes of
frequency $\omega$ are related by $(\delta T/T)_{\rm core} \sim
(dV_{\rm core}/dr)/\omega \sim V_{\rm core} /(\omega R_{\rm core})$.
Thus,
$V_{\rm core} \sim (\delta T/T)_{\rm core} R_{\rm core}\omega_{\rm
core}
\sim 10^7 (\delta T/T)_{\rm core}$ cm s$^{-1}$; these relations are
derived by making use of the linearized mass and entropy equations.

\bigskip
\noindent
\centerline{3.\  EIGENMODES AND EIGENVELOCITIES}
\medskip

We have calculated illustrative g-mode eigenfunctions and frequencies
for a standard solar model by numerically integrating adiabatic wave
equations which includes the perturbation to the gravitational
potential
(the relevant equations and boundary conditions are described in
Christensen-Dalsgaard \& Berthomieu 1991 and in Kumar et al. 1992).

Table~1 gives the calculated characteristics of some typical g-modes
with
relatively small number of nodes.  The Table presents, as a function
of  the radial node, $n$, and the angular degree, $\ell$, the period,
the mode energy, and the amplitude of the temperature variation. We
have normalized the mode energy and the temperature fluctuation to a
RMS surface velocity of 1 cm/s (averaged over spherical surface and
time). For this normalization, the typical temperature variations in
the solar core, $\delta T/T$, are
between $10^{-7}$ and $10^{-6}$.  The energy in each mode scales
approximately as $ (\delta T/T)^2$. If we require that the temperature
fluctuation be large enough to change the $^8$B solar neutrino flux, $
(\delta T/T)~=~0.1$, we obtain mode energies (see the table) in
agreement
with those estimated in the previous section, $\sim 10^{43}$ erg.

\topinsert
$$\vbox{\baselineskip=1.5\normalbaselineskip
\halign{\hfil$#$\hfil\tabskip=3em plus2em
minus2em&\hfil$#$\hfil&\hfil$#$&\hfil$#$&\hfil$#$&\hfil$#$\tabskip=0pt\cr
\multispan6{\hfil Table 1.\ Some Characteristics of Typical Low-order
Solar
g-modes.\hfil}\cr
\noalign{\medskip\hrule\smallskip\hrule\medskip}
n&l&\hfil \nu\hfil&\hfil{\rm Period}\hfil&\hfil{\rm
Energy}^\dagger\hfil&\hfil
\delta T/T^\dagger\hfil\cr
&&\hfil (\mu Hz)\hfil&\hfil {\rm (hr)}\hfil&\hfil{\rm(erg)}\hfil\cr
\noalign{\medskip\hrule\medskip}
1&1&267.4&1.04&3.4\times 10^{31}&6.7\times 10^{-8}\cr
1&2&297.1&0.94&2.1\times 10^{31}&4.4\times 10^{-8}\cr
1&3&340.2&0.82&4.3\times 10^{31}&7.7\times 10^{-8}\cr
2&1&193.0&1.44&1.3\times 10^{33}&4.6\times 10^{-7}\cr
2&2&259.7&1.07&3.7\times 10^{31}&7.2\times 10^{-8}\cr
2&3&299.2&0.93&4.4\times 10^{31}&8.8\times 10^{-8}\cr
3&1&153.7&1.81&1.7\times 10^{33}&6.9\times 10^{-7}\cr
3&2&224.0&1.24&8.7\times 10^{31}&1.3\times 10^{-7}\cr
3&3&264.7&1.05&3.3\times 10^{31}&8.1\times 10^{-8}\cr
4&1&128.3&2.17&1.7\times 10^{33}&7.4\times 10^{-7}\cr
4&2&195.5&1.42&1.6\times 10^{32}&2.0\times 10^{-7}\cr
4&3&240.5&1.15&2.9\times 10^{31}&7.4\times 10^{-8}\cr
5&1&109.9&2.53&1.7\times 10^{33}&8.7\times 10^{-7}\cr
5&2&171.8&1.62&3.1\times 10^{32}&3.1\times 10^{-7}\cr
5&3&218.9&1.27&6.8\times 10^{31}&1.3\times 10^{-7}\cr
\noalign{\medskip\hrule\medskip}
\multispan6{$^\dagger$\ Normalized to a velocity of 1 cm/s at the solar
surface.\hfill}\cr
}\smallskip}
$$
\endinsert

The characteristic velocities at the surface scale proportionally to
the hypothesized amplitude of the oscillation, i.e., proportional to $
(\delta T/T)$, and hence can be estimated directly from Table~1. We
find
that the surface velocities for the low degree g-modes with $ (\delta
T/T)~=~0.1$ are typically a km/s, approximately $10^4$ times larger
than the typical observed limits  of 10 cm/s (see Kuhn, Libbrecht, \&
Dicke 1986). According to our numerical calculations, g-modes of
degree
less than about 15 can not simultaneously satisfy both the large
energy
requirement for the solution to the neutrino problem as well as the
observational upper limit on the surface velocity amplitude. From
this we
conclude that g-modes of degree less than 15, with large enough
amplitude to solve the neutrino problem, are ruled out by current
observations.

One might try to avoid this contradiction with observation by
proposing
that the g-modes are damped by dissipative processes that do not occur
in the core but instead occur outside the region in which the
neutrinos are
produced. However, the damping required to decrease the large surface
amplitude to a value below the observational limit is so strong that
it will kill the instability responsible for mode excitation.

Another possible way to hide large amplitude g-modes in the solar
core is to
appeal to modes of degree greater than approximately 15. As discussed
earlier, high degree g-modes, with $\delta$T/T of 0.1 in the core,
will
have surface velocities that are below the observational limits.
However, there are several problems with this proposal.  First, no one
finds that these high-$\ell$ modes are overstable; so one needs to
appeal to an unknown mechanism for their excitation.  Moreover, the
mechanism that excites $\ell$=15 modes is very likely to excite modes
of lower degree as well to similar energies; and the energies of lower
degree modes are subject to the tight observational constraints.
Also,
the requirements on the total mode energy and on the temperature
fluctuation required to change the neutrino fluxes are more stringent
for higher degree modes than for the previously discussed
case of a single, low-order
g-mode.

Transferring the energy from low $\ell$ modes to high $\ell$ modes
will
also not work.  The rate at which energy is injected into the high
$\ell$ modes would be approximately equal to the total energy in low
degree modes, $E_{{\rm low}-\ell}$ divided by their characteristic
growth time, $\tau_{{\rm low}-\ell}$, of order a few million years
(see
Saio 1980). It is easy to see that the total energy in the high $\ell$
modes must be less than ($\tau_{ {\rm high}-\ell }/\tau_{ {\rm
low}-\ell})\times E_{ {\rm low}-\ell}$.  The characteristic
dissipation
time for the high $\ell$ modes, $\tau_{{\rm high}-\ell}$, is $\sim
10^{5.5}$ y (Shibahashi et al. 1975).  Therefore, the total energy in
the high $\ell$ modes is expected to be an order of magnitude less
than
the (inadequate) total energy in the low degree modes.  As we have
seen
earlier, the observational upper limits on the surface velocities of
low degree g-modes require that their energies be small, 10$^{-8}$
times the energy needed to solve the neutrino problem.

\bigskip
\noindent
\centerline{4.  DISCUSSION AND CONCLUSIONS}
\medskip
\medskip
\noindent

The order-of-magnitude considerations discussed in \S\ 2 apply to any
hypothesized large-amplitude, short-period oscillations that grow on
a Kelvin-Helmholtz time scale
in the solar core and that one might consider as a possible solution
of the
solar neutrino problem.  In particular, such solutions
require large amounts of
energy, \eqt{\ratiogtop} (for $g$-modes, $10^9$ times the energy in
the
observed $p$-modes), and a tiny amount of dissipation, \eqt{\QRatio}
(for $g-$modes, $10^{-8}$ the fractional dissipation rate of the
$p$-modes).
We find that for low degree $g_1$ modes the turbulence dissipation
rate
is larger by a factor of about 40 than the growth rate due
to the $^3$He instability. The uncertainty in the calculation of
the turbulence dissipation rate is unlikely to be as large as a
factor of
40, and so we believe that low degree $g_1$ modes are stabilized by
turbulence dissipation.

By explicit calculation, we
show in \S\ 3 that low order g-mode oscillations with large enough
amplitudes to affect the calculated solar neutrino fluxes would
produce
large periodic velocity shifts at the solar surface that are not
observed.  We conclude that it is unlikely that the solution of the
solar neutrino problem depends upon large amplitude oscillations of
the
kind explored here.

\bigskip
\medskip
We are grateful to J. De R\'ujula, S. Glashow, P. Goldreich, W. Press,
J. Toomre, and R. Ulrich for valuable discussions and suggestions.
PK is grateful to Paul Schechter for encouragement and advice.
This work was supported in part by NSF grant no.~PHY-92-45317 with the
Institute for Advanced Study and by NASA contract W-17677 with MIT.
\bigskip
\medskip
\centerline{REFERENCES}

\Rf
Bahcall, J.N., \& Ulrich, R. K. 1988, Rev. Mod. Phys., 60, 297
\Rf
Boury, A., Gabriel, M., Noels, A., Scuflaire, R., \& Ledoux, P.
1975, AA,
41, 279
\Rf
Brown, T. M., Mihalas, B. W., \& Rhodes, E. J. Jr. 1986, in Physics of
the Sun, Vol. I: The Solar Interior, ed. P. A. Sturrock, T. E. Holzer,
D. M. Mihalas, \& R. K. Ulrich (Dordrecht: Reidel), 171
\Rf
Christensen-Dalsgaard, J., \& Berthomieu, G. 1991, in Solar Interior
and
Atmosphere, ed. A. N. Cox, W. C. Livingston, \& M. Matthews (Tucson:
Univ. Arizona Press), in press
\Rf
Christensen-Dalsgaard, J., Dilke, J. F. W. W., \& Gough, D. O. 1974,
M.N.R.A.S., 169, 429
\Rf
De R\'ujula, A., \& Glashow, S. 1992, CERN-Th 6608/92 preprint (to be
published)
\Rf
Dilke, J. F. W. W., \& Gough, D. O. 1972, Nature, 240, 262
\Rf
Goldreich, P., \& Keeley, D.K. 1977, ApJ, 212, 243
\Rf
Goldreich, P., \& Kumar, P. 1988, ApJ, 326, 462
\Rf
Gough, D. O. 1991, Annals of the New York Academy of Sciences, 647,
199
\Rf
Guzik, J. A., \& Cox, A. N. 1992, ApJ, 386, 729
\Rf
Harvey, J. W. 1990, in Progress of Seismology of the Sun and Stars,
ed.
Y. Osaki, \& H. Shibahashi, Lec. Notes Phys., vol. 367 (Heidelberg:
Springer), 115
\Rf
Kuhn, J. R., Libbrecht, K. G., \& Dicke, R. H. 1986, Nature, 319, 128
\Rf
Kumar, P., Goldreich, P., \& Kerswell, R. 1992, in preparation
\Rf
Landau, L. D. \& Lifshitz, E. M. 1982, Fluid Mechanics, Pergamon press
\Rf
Leibacher, J. W., Noyes, R. W., Toomre, J., \& Ulrich, R. K. 1985,
Scientific Amer., 253(3), 48
\Rf
Libbrecht, K. G., Popp, B. D., Kaufman, J. M., \& Penn, M. J. 1986,
Nature, 323, 235
\Rf
Libbrecht, K. G., \& Woodard, M. F. 1991, Science, 253, 152
\Rf
Merryfield, W. J., Toomre, J., \& Gough, D. O. 1991, ApJ, 367, 658
\Rf
Press, W. H. 1981, ApJ, 245, 286
\Rf
Press, W. H., \& Rybicki, G. B. 1981, ApJ, 248, 751
\Rf
Press, W. H. 1986, in Physics of
the Sun, Vol. I: The Solar Interior, ed. P. A. Sturrock, T. E. Holzer,
D. M. Mihalas, \& R. K. Ulrich (Dordrecht: Reidel), 77
\Rf
Rosenbluth, M., \& Bahcall, J. N.  1973, ApJ, 184, 9
\Rf
Saio, H. 1980, ApJ, 240, 685
\Rf
Shibahashi, H., Osaki, Y., \& Unno, W. 1975, Publ. Astron. Soc.
Japan, 27, 401
\Rf
Spruit, H. C. 1987, in The Internal Solar Angular Velocity, ed. B. R.
Durney, \& S. Sofia (Dordrecht: Reidel), 185
\Rf
Toomre, J. 1986, in  Seismology of the Sun and the Distant Stars,
ed. D. O. Gough, NATO ASIC169 (Dordrecht: Reidel), 1
\Rf
Ulrich, R. K. 1974, ApJ 188, 369
\end